\documentclass[12pt]{article}
\usepackage{cite,graphicx,amsmath,amssymb}
\usepackage{graphicx}
\usepackage{multicol}
\usepackage{xcolor}
\usepackage{slashed}
\usepackage[normalem]{ulem}

\newcommand{\be}{\begin{equation}}
\newcommand{\ee}{\end{equation}}
\newcommand{\bea}{\begin{eqnarray}}
\newcommand{\eea}{\end{eqnarray}}
\begin{document}

\thispagestyle{empty}
\begin{flushright}
CCTP-2018-15, ITCP-IPP 2018/11, MAD-TH-18-08
\end{flushright}

\vfil
\vspace*{-.2cm}
\noindent

\vspace*{1.2cm}

\begin{center}

{\Large\bf Brane gaugino condensate in 10d}
\\[2cm]

{\large Yuta Hamada$^{1}$, Arthur Hebecker$^{2}$, Gary Shiu$^{3}$ and Pablo Soler$^{2}$}

\vfil

{\it

${}^{1}$ {Crete Center for Theoretical Physics, Institute for Theoretical and Computational Physics, Department of Physics, University of Crete, P.O. Box 2208, 71003 Heraklion, Greece}\\
 ${}^{2}$ Institute for Theoretical Physics, University of Heidelberg, 
Philosophenweg 19,\\ D-69120 Heidelberg, Germany\\
${}^{3}$ Department of Physics, University of Wisconsin, Madison, WI 53706, USA
}

\large{December 14, 2018}
\\[1.6cm]

{\bf Abstract}
\end{center} 

We analyze the structure of gaugino interactions on D7-branes from a 10d perspective. This is essential if one wants to lift the standard 4d approach to type IIB moduli stabilization to 10d. In particular, a 10d picture has recently been used to raise concerns about the KKLT proposal for constructing de Sitter vacua, and to lend support to swampland conjectures against de Sitter. However, the analyses of brane gaugino condensation so far are plagued by UV divergences and do not include local 4-fermion terms. They also fail to reproduce the 4-fermion terms required by supergravity when compactified to four dimensions. Motivated by the structure of heterotic and Ho\v rava-Witten theories, we suggest an extension of the brane action by a particular 4-fermion operator that resolves the above problems. Crucially, the UV divergence is cancelled and the expected structure of the 4d effective action is reproduced. We believe that attempts at a 10d description of KKLT have to be reconsidered in this new light.

\newpage

\section{Introduction}
Our current understanding of string theory suggests that to make contact with the real world, the moduli describing the sizes and shapes of compactifications should be stabilized
at values that give physically acceptable 4d couplings. 
The observational evidence for dark energy adds another layer of complication: not only are the moduli
stabilized but in such a way that leads to a positive vacuum energy.
The simplest option to accommodate the late-time cosmological data is to consider
stabilizing moduli in a de Sitter vacuum.
By now, a variety of seminal frameworks for constructing de Sitter vacua in string theory have been proposed, most prominently the KKLT~\cite{Kachru:2003aw} and LVS~\cite{Balasubramanian:2005zx} scenarios (for earlier supercritical constructions, see~\cite{Silverstein:2001xn,Maloney:2002rr}). Due to the complexity of these scenarios, 
it is fair to say that the constructions so far are not yet fully explicit. 
Much of the discussions are at the level of proposed scenarios rather than detailed models. 
At the same time, concerns have been raised regarding the reliability of these constructions. For arguments pro and con see, e.g.~\cite{Bena:2009xk,McOrist:2012yc,Bena:2014jaa,Cohen-Maldonado:2015ssa,Junghans:2016abx,Moritz:2017xto,Sethi:2017phn,Danielsson:2018ztv,Moritz:2018sui,Cicoli:2018kdo,Kachru:2018aqn,Kallosh:2018nrk,Akrami:2018ylq,Moritz:2018ani,Bena:2018fqc,Kallosh:2018psh,Gautason:2018gln,Armas:2018rsy}.
This state of affairs has motivated the conjecture
that metastable de Sitter vacua do not arise in parametrically controlled regimes of string theory~\cite{Obied:2018sgi}, with supporting first-principles arguments discussed in~\cite{Ooguri:2018wrx} (see also~\cite{Hebecker:2018vxz}).\footnote{
The necessary refinements~\cite{Ooguri:2018wrx} (see also~\cite{Garg:2018reu}) of the original de Sitter conjecture~\cite{Obied:2018sgi} are not essential in our context.
}
This conjecture, if true, would imply that the observed dark energy is time dependent, though concrete models utilizing this option
have yet to be constructed.

In light of the above considerations, it is clearly of utmost importance to 
firm up the proposed constructions of de Sitter vacua in string theory. In this paper, we focus on the KKLT and LVS scenarios as they share the common feature of stabilizing moduli with non-perturbative effects such as gaugino condensation.
Since these constructions involve 4d ingredients, a way to address the concern of whether there is a separation of scales is to lift the 
original 4d effective field theory analysis to 10d. 
This approach has led to concerns~\cite{Moritz:2017xto} which have been further discussed in~\cite{Gautason:2018gln}. A key point in these analyses is the 10d description of the gaugino condensate on a D7 brane stack and its coupling to bulk fluxes~\cite{Camara:2004jj,Baumann:2006th,Koerber:2007xk,Baumann:2009qx,Baumann:2010sx,Dymarsky:2010mf}. As emphasized in~\cite{Gautason:2018gln}, the precise form of the localized brane effects is crucial for making the Maldacena-Nunez-type no-go argument~\cite{Maldacena:2000mw} suggested in~\cite{Moritz:2017xto}. This is one of our motivations for finding a precise 10d description of brane gaugino interactions. 
Another is the desire, independent of specific model building challenges, to take the understanding of similar dynamics in the heterotic~\cite{Derendinger:1985kk,Dine:1985rz} and M-theory context \cite{Horava:1995qa,Horava:1996ma,Horava:1996vs}  (see also~\cite{Nilles:1997cm,Quigley:2015jia}) to the realm of type IIB string theory.

Our analysis suggests that two important effects have so far been missed: First, there are 4-fermion operators on the brane or, equivalently, a counterterm subtracting local divergences~\cite{Michel:2014lva,Polchinski:2015bea}. Second, compactification or flux-quantization effects turn out to be essential for the dynamics of the bulk fluxes sourced by the gaugino condensate. 

As will be discussed in detail later, the need for a counterterm arises since, symbolically, the 10d action takes the form
\be
\int \left[\,\frac{1}{2}G(x)^2+G(x)j(x)\, \right]\qquad\mbox{with}\qquad
j(x)\sim \langle \lambda^2 \rangle \delta(x)\,.\label{nsq}
\ee
where $\delta(x)$ represents the localization of the source to a certain submanifold.
The bulk flux $G$ then develops a profile including a singular piece proportional to the gaugino condensate, $G\sim \langle \lambda^2 \rangle \delta(x)$. Inserting this in the action leads to a divergence of the type $\delta(0)$. 
In fact, a similar divergence appears in Ho\v rava-Witten theory~\cite{Horava:1995qa,Horava:1996ma,Horava:1996vs,Mirabelli:1997aj}, where it was argued using supersymmetry that a quartic gaugino term cancelling the divergence must be present. Moreover, this term leads to a perfect square structure of the action. Motivated by this, we propose a choice of counterterm realizing such a structure also in the D7 brane case:
\be
\int\frac{1}{2}\Big[G(x)+\tilde{\jmath}(x)\Big]^2\,.\label{tsq}
\ee 
Our proposal is supported by finiteness, by the resulting gaugino terms in 4d effective supergravity, and by dualities with heterotic and Ho\v rava-Witten theories. In addition, it reproduces the structure of the scalar potential expected from 4d considerations.

Compactification effects are also essential for the final result. Indeed, the singular source has been modified in going from $j$ in (\ref{nsq}) to $\tilde{\jmath}$ in (\ref{tsq}) in order to allow $G$ to cancel it precisely. That is, in a non-compact geometry $G$ backreacts such that the total square vanishes. By contrast, in a compact geometry flux quantisation makes this impossible. The total energy is then finite and bulk-dominated.
We note that our modifications are not in conflict with the analyses of~\cite{Baumann:2006th,Baumann:2010sx}. The motivation of these papers was the effect of the flux, sourced by the gaugino condensate, on  D3-branes at a distant locus. In this case, the brane-localised singularities and compactification effects are not essential. However, if one is interested in the total vacuum energy of a given compactification, these effects are of primary importance.

Our results indicate that the analyses of~\cite{Moritz:2017xto,Gautason:2018gln} should be revisited, which we leave to future work. Currently, we do not find an immediate obstruction to possible uplifts to de Sitter upon inclusion of anti-branes. 

This paper is organized as follows. In Sec.~\ref{Sec:co-dimension_one}, the treatment of divergent terms is explained along the lines of Ho\v rava-Witten theory, using in particular the toy model approach of \cite{Mirabelli:1997aj}. In Sec.~\ref{Sec:co-dimension_two}, we propose an action with a perfect square structure suitable for the $D7$ brane case. We summarize our findings and conclude in Sec.~\ref{Sec:conclusion}.

\section{Localized sources in co-dimension one}\label{Sec:co-dimension_one}
In this section, we first review how the divergent terms are treated in Ho\v rava-Witten theory. Then, as an exercise for next section, we discuss the effective action after integrating out the flux in a five dimensional toy model.

Bulk fluxes coupled to localized sources, and their associated divergences have been studied in string theory for a long time. They were first discussed in Ho\v rava-Witten theory~\cite{Horava:1995qa,Horava:1996ma,Horava:1996vs}, i.e. M-theory compactified on an interval $M^{11}=M^{10}\times I$, which is dual to strongly coupled $E_8\times E_8$ heterotic theory. In this setup, a bulk 4-form field strength $G$ is sourced by gaugino condensation in the $E_8$ gauge theory living on one of the (real) co-dimension-one boundaries. 

The relevant terms of the action that involve the 4-form flux are the bulk kinetic term and the localized coupling to the source $j_{ABC}\sim \langle \bar{\chi}^a\Gamma_{ABC} \chi^a\rangle$. We assume that the source is constant in 10d, and use $A,B,\ldots$ for 10-dimensional indices tangent to the boundary, and $I,J,\ldots$ for 11-dimensional ones. Schematically, the action contains:
\be\label{eq:SG}
S_G\sim -\int_{M^{11}}d^{11}x\,\sqrt{-g} \left(G_{IJKL}G^{IJKL}-\delta(x^{11})G_{ABC11}j^{ABC}\right).
\ee
Here we have placed the boundary where gaugino condensation occurs at $x_{11}=0$. At this point we only want to describe the structure of the action, and hence are not keeping track of overall coefficients, or even relative ones (the precise action can be found in~\cite{Horava:1996ma}). 

It is immediate from~\eqref{eq:SG} that the solutions to the equations of motion involve fluxes that diverge at the boundary $G_{ABC11}\sim \delta(x^{11})j_{ABC}$. The $\delta$-function should be interpreted as a UV-divergence that would be  regularized in a full microscopic treatment where the boundary has a finite thickness. What is worrisome, however, is that the action~\eqref{eq:SG} is UV divergent on-shell, $S_G\sim \delta(0)$, and hence sensitive to the regularization procedure. 

The fate of this infinite contribution was clarified in~\cite{Horava:1996ma,Horava:1996vs,Mirabelli:1997aj}. It was shown that supersymmetry requires the presence of a quartic gaugino  term in the action that precisely cancels the divergence. In fact, adding this term to~\eqref{eq:SG}, the action involving $G_{ABC11}$ takes a simple quadratic form
\be\label{eq:HW}
S\sim - \int_{M^{11}}d^{11}x\,\sqrt{-g} \left(G_{ABC11}-\frac{1}{2}\delta(x^{11})j_{ABC}\right)^2.
\ee
The quadratic form of the action is reminiscent of the one that arises in weakly coupled heterotic theory~\cite{Dine:1985rz} and can be expected from duality considerations to hold rather generally. The difference with the heterotic case is that the gaugino source is localized to a boundary while the flux propagates in the bulk, leading to the mentioned divergences.

It is clear, nevertheless, that the on-shell action~\eqref{eq:HW} will be finite and will in fact vanish. That is, unless there is a topological obstruction that prevents the $G$-flux to exactly compensate for the gaugino contribution.

In order to see how this works in more detail, we discuss now a greatly simplified toy model. We consider a five dimensional field theory of a zero-form gauge field $\phi$. This field couples to a constant source $j$ that is localized in the fifth dimension, at $y=0$. In analogy to~\eqref{eq:HW}, we take the action to be 
\be\label{eq:1dtoy}
S= - \int_{M^{5}} \left(d\phi-j\delta(y)dy\right)\wedge \ast \left(d\phi-j\delta(y)dy\right).
\ee
A supersymmetric version of this model was considered in~\cite{Mirabelli:1997aj}, where the quadratic structure of the action was shown to be required by supersymmetry.

The equation of motion for $\phi$ reads
\be
d\ast\left(d\phi-j\delta(y)dy\right)=0\qquad \Longrightarrow\qquad d\phi=j\delta(y)dy + \alpha_{I}dx^{I}.
\ee
where $\alpha$ is a co-closed one-form, $d\ast \alpha=0$. The Bianchi identity $d^2\phi=0$ implies that $\alpha$ is also closed, and hence harmonic. If $\alpha$ can be set to zero, the action vanishes trivially on-shell, despite the divergent flux. 

There may be an obstruction, however, if the fifth dimension admits a non-trivial one cycle. Let us consider the case where the fifth dimension is a circle. In this case, flux quantization on the circle implies
\be
\int_{S^1}d\phi\in {\mathbb Z} \qquad \Longrightarrow \qquad \int dy \, \partial_y\phi = j+\alpha_5 = n \in {\mathbb Z} \qquad \Longrightarrow \qquad  \alpha_5 = n -j. 
\ee
where we have taken the circle to have unit length. One sees that for a generic source $j$ the action does not vanish but acquires a finite value
\be
S=- \int_{M^{4}\times S^{1}} \left(n-j\right)dy\wedge \ast \left(n-j\right)dy=-(n-j)^2.
\ee
We can take $n=0$ if we wish, but we are always left with a finite action proportional to $j^2$. Recall that this source is the analog of the gaugino bilinear that condenses in the string theory models we seek to understand. The {\it finite} contribution we just found will generalize to those cases naturally.

Before going to the more relevant case of co-dimension-two branes, we would like to describe a generalization of the model just discussed, by including two different sources localized in different points $y_1$ and $y_2$ of the internal circle. The action is simply
\be
S= -\int_{M^{5}} \left[d\phi-j_1\delta(y-y_1)dy-j_2\delta(y-y_2)dy\right]\wedge \ast \left[d\phi-j_1\delta(y-y_1)dy-j_2\delta(y-y_2)dy\right].
\ee
Just like before, the solutions satisfy
\be
d\phi= \left[j_1\delta(y-y_1)+j_2\delta(y-y_2)\right]dy + (n-j_1-j_2)dy.
\ee
where flux quantization has already been imposed by $n\in{\mathbb Z}$. The total on-shell action is now
\be\label{eq:separated}
S=-(j_1+j_2-n)^2.
\ee
The cross term $j_1 j_2$ will have some relevance in our later discussion. It corresponds to a quartic gaugino interaction between gaugino bilinears on {\it different}, separated branes.

\section{Localized sources in co-dimension two}\label{Sec:co-dimension_two}
We discuss in this section the coupling of gauge fields to sources localized on (real) co-dimension-two submanifolds. Gaugino condensation on D7-branes provides sources of this type. Before going to that system, in Sec.~\ref{Sec:toy_model}, we consider a toy model which is a generalization of the previous setup~\eqref{eq:1dtoy} to six dimensions. After that we move on to the realistic setup of D7-branes in type IIB string theory in Sec.~\ref{Sec:type_IIB}.

\subsection{A toy model}\label{Sec:toy_model}
\subsubsection{Setup and problem}
For the moment, we work in non-compact flat space ${\mathbb R}^6={\mathbb R}^4\times {\mathbb C}$, with ${\mathbb C}$ parameterized by a complex coordinate $z$. The source is localized at $z=0$. This is described by a delta-function two-form
\be
\delta^{(2)}=\frac{1}{\pi}\partial\bar{\partial}\log|z|,
\ee
where we have introduced Dolbeault (exterior derivative) operators $\partial$ and $\bar{\partial}$:
\be
\partial\equiv dz \frac{\partial}{\partial z}\wedge\,,\qquad 
\bar{\partial}\equiv d\bar{z} \frac{\partial}{\partial \bar{z}}\wedge.
\ee
From this, one can obtain the scalar delta-function
\be\label{eq:delta}
\delta^{(0)}=g^{\alpha\bar{\beta}}\delta^{(2)}_{\alpha\bar{\beta}}
=\frac{2}{\pi}\partial_z\bar{\partial}_z\log|z|,
\ee
normalized by $\int dxdy\, \delta^{(0)}=1$ for $z=x+iy$.
 We use subscripts to distinguish partial derivatives $\partial_z$ and $\bar{\partial}_{z}$ from Dolbeault operators $\partial$ and $\bar{\partial}$:
\be
\partial_z\equiv\frac{\partial}{\partial z}\,,\qquad 
\bar{\partial}_{z}\equiv\frac{\partial}{\partial \bar{z}}.
\ee
We want to consider the coupling of the one-form field strength $G_1=d\phi$ to a localized holomorphic current $j_1=jdz$, where $j$ is constant: 
\be\label{eq:Gj}
{\cal L}_{jG}\sim  (G_1\cdot j_1)\Big{|}_{\{z=0\}}= \int_{{\mathbb C}}(G_1\cdot j_1) \,\delta^{(2)}=\int_{{\mathbb C}}G_1\wedge  j_1\, \delta^{(0)}.
\ee
Here we think of ${\cal L}$ as of a 4d lagrangian obtained after integrating over the extra (at the moment non-compact) dimensions. By integrating out $G_1$, we encounter the divergence of the on-shell action, as in the co-dimension one case.

In analogy to the co-dimension-one example, one might expect that we can remove the divergence if we write an action with a perfect square structure that contains the flux kinetic term as well as its coupling to the source:\footnote{Here and in what follows we focus only on the dependence on the directions parameterized by $z$. In particular, the Hodge $\ast$-operator corresponds to the two-dimensional one $\ast_2$. We will often make use of the imaginary (anti-) self duality properties $\ast dz = -i dz$ and $\ast d\bar{z}= id\bar{z}$.}
\begin{eqnarray}\label{eq:2dtoy}
{\cal L}&=& -\int_{{\mathbb C}} \left(G_1-\overline{j}_1\delta^{(0)}\right)\wedge \ast \left(\overline{G}_1- j_1\delta^{(0)}\right)\nonumber\\
&=& -\left(G_1-\overline{j}_1\delta^{(0)},G_1-\overline{j}_1\delta^{(0)}\right)=-\Big|G_1-\overline{j}_1\delta^{(0)}\Big|^2,
\end{eqnarray} 
where we have introduced the inner product
\be
(A,B)=\int_{\mathbb C}A\wedge \ast \overline{B}\,,\qquad (A,A)=|A|^2.
\ee

The Bianchi identity and the equations of motion from~\eqref{eq:2dtoy} are simply
\be\label{eq:2deqs}
dG_1=0\,,\qquad d\ast \left(G_1-\overline{j}_1\delta^{(0)}\right)=0.
\ee
However, we now encounter a problem that was not present in the co-dimension one case. Even if we are working in non-compact space, we cannot simultaneously solve the equations of motion and the Bianchi identities, and obtain a vanishing (or even finite) action. The reason is that the source term in the equation of motion $\overline{j}_1\delta^{(0)}$ is not closed (because of the delta-function), and hence one cannot set $G_1=\overline{j}_1\delta^{(0)}$. This would be incompatible with the Bianchi identity. 

To see the problem in more detail, express the flux in terms of its holomorphic and anti-holomorphic components
\be
G_1=g \,dz + \tilde{g}\,d\bar{z}.
\ee
Using that $dz$ is imaginary anti-self-dual (IASD), $\ast dz = -i dz$; while $d\bar{z}$ is imaginary self-dual (ISD), $\ast d\bar{z}= id\bar{z}$; equations~\eqref{eq:2deqs} can be combined into
\be\label{eq:componenteqs}
\bar{\partial}_{z}g=\partial_z\tilde{g}=\frac{\bar{j}}{2}  \partial_z\delta^{(0)}(z).
\ee
One can give an explicit solution to~\eqref{eq:componenteqs} with the help of the delta-function expression~\eqref{eq:delta}:
\be
g=\frac{\bar{j}}{\pi}\partial_z\partial_z \ln|z|\,\qquad \tilde{g}=\frac{\bar{j}}{\pi}\partial_z\bar{\partial}_{z} \ln|z|=\frac{\bar{j}}{2}\delta^{(0)}(z),
\ee
which corresponds to a total flux\footnote{We are always free to add a harmonic form to this solution. We will do so momentarily.}
\be\label{eq:solG1}
G_1=g dz+\tilde{g}d\bar{z}= \frac{\bar{j}}{\pi}\left(\partial+\bar{\partial}\right)(\partial_z\ln|z|)= \frac{\bar{j}}{\pi}d(\partial_z\ln|z|).
\ee
One can then see that the action will be divergent, since it is the square of a divergent form:
\be\label{eq:coexactpiece}
G_1-\overline{j}_1\delta^{(0)}=\frac{\bar{j}}{\pi}\left(\partial-\bar\partial\right)(\partial_z\ln|z|)\qquad \Longrightarrow\qquad S\sim \delta(0).
\ee
This divergence is a strong indication that the action we started with, eq.~\eqref{eq:2dtoy}, is not the correct one. In particular, the constant divergent piece $\left(j\delta^{(0)},j\delta^{(0)}\right)$ is not appropriately absorbing the divergence from the flux. It would seem that simply adding a divergent term to cancel this divergence would spoil the nice perfect square structure that one expects from heterotic and Ho\v rava-Witten theory. We show in the following that this is not the case.

\subsubsection{The finite action}

The resolution of the problem above is not hard to find. As we mentioned, the obstruction arises from the fact that the source contribution $\overline{j}_1\delta^{(0)}$ is not closed, and hence cannot be compensated by a flux that satisfies the Bianchi identity. We would hence like to project this term onto the subspace of closed forms, but without modifying the equations of motion for $G_1$ and the perfect square structure. 

One can easily see that this is always possible. In a compact space without boundary, which is ultimately the setup we want to consider, any form can be written as a closed piece (harmonic plus exact) plus a co-exact one:
\be\label{eq:hodgedecomposition}
\omega= \alpha + d\beta + d^{\dagger}\gamma,
\ee
where $\alpha$ is harmonic and $d^{\dagger}=\ast d \ast$. So what we need is to subtract the co-exact component of the source $\overline{j}_1\delta^{(0)}$. This procedure, in turn, does not modify the equations of motion for $G$, since the inner product of a closed form (such as $G_1$) with a co-exact form vanishes:
\be\label{eq:inner}
(G_1,d^{\dagger}\beta)=(dG_1,\beta)=0.
\ee
Hence, one can shift  the source with a co-exact term without affecting the equations of motion of $G$ nor its solutions. 

This argument also holds in the non-compact model of the previous subsection. It can be easily seen that the source contribution decomposes as in~\eqref{eq:hodgedecomposition}:
\begin{eqnarray}
\bar{j}_1\delta^{(0)}(z)&=&\frac{2\bar{j}}{\pi}\bar\partial(\partial_z \log|z|)\nonumber\\
&=&\frac{\bar{j}}{\pi}(\partial+\bar{\partial})(\partial_z \log|z|)
-\frac{\bar{j}}{\pi}(\partial-\bar{\partial})(\partial_z \log|z|)\nonumber\\
&=&\frac{\bar{j}}{\pi}d(\partial_z \log|z|)
+\frac{\bar{j}}{2\pi}d^{\dagger}[(\partial_z \log|z|)dz\wedge d\bar z].
\end{eqnarray}
The co-exact piece is precisely the term~\eqref{eq:coexactpiece} that could not be balanced by flux. What we know now is that this piece can be simply dropped from the action without otherwise spoiling its perfect square structure nor changing the equations of motion:\footnote{In a non-compact space eq.~\eqref{eq:inner} holds only up to a boundary term, but this does not contribute to the equations of motion.}
\begin{eqnarray}\label{eq:finiteactiononeform}
{\cal L} &\to&  -\int_{{\mathbb C}} \left(G_1-\frac{\bar{j}}{\pi}d(\partial_z \log|z|)\right)\wedge \ast \left(\overline{G}_1-\frac{j}{\pi}d(\bar{\partial}_z \log|z|)\right)\nonumber\\
&&=-\Big|G_1-\frac{\bar{j}}{\pi}d(\partial_z \log|z|)\Big|^2.
\end{eqnarray}

The solution to the equations of motion is still~\eqref{eq:solG1}, to which we can add an arbitrary harmonic piece $G^{(0)}_1$:
\be
G_1=\frac{\bar{j}}{\pi}d(\partial_z\ln|z|)+G_1^{(0)}.
\ee
Plugging this back into the action~\eqref{eq:finiteactiononeform} one immediately gets the finite result
\be
{\cal L}=-(G_1^{(0)},G_1^{(0)}).
\ee
In a non-compact space, there is no obstruction to setting $G_1^{(0)}=0$. The flux can exactly compensate the contribution from the source to yield a vanishing action. This suggests that the projection performed to reach~\eqref{eq:finiteactiononeform} is the correct prescription. It leads to the correct action of a bulk flux coupled to a source localized to a co-dimension two divisor.

\subsubsection{Sources in a compact space}
As in the previous section, we now consider the constraints imposed by flux quantization when the source is localized on a compact space. For simplicity, we take this space to be a square torus of area $A_{T^2}=L^2$, parametrized by $z\sim z+L\sim z+ iL$. 

The first difference with respect to the non-compact case is that the Green function on a torus is no longer given as in~\eqref{eq:delta} by the logarithm because this would not be globally well-defined. Instead, it satisfies
\be
4\partial_z\bar{\partial}_zG_{T^2}(z)
=\delta^{(0)}-\frac{1}{A_{T^2}}.
\ee
The necessity of the constant factor can be seen by integrating both sides of this equation. $G_{T^2}$ can be expressed in terms of elliptic $\vartheta$-functions, but we will not need its explicit form here.

As before, the action takes a quadratic form
\begin{eqnarray}
{\cal L} &=& - \int_{T^2} \left(G_1-\overline{j}_1\delta^{(0)}\right)\wedge \ast \left(\overline{G}_1-j_1\delta^{(0)}\right)\\
&\to& - \int_{T^2} \left(G_1-2\bar{j}d(\partial_z G_{T^2})-\frac{\bar{j}}{A_{T^2}}d\bar z\right)\wedge \ast\left(\overline{G}_1-2 j d(\bar{\partial}_z G_{T^2})-\frac{j}{A_{T^2}}dz\right),\nonumber
\end{eqnarray}
where in the second line we have projected the source onto the sub-space of closed forms. The solution is
\be
G_1 = 2\bar{j}d(\partial_z G_{T^2}) + G_1^{(0)},
\ee
where again, $G_1^{(0)}$ is a harmonic (i.e. constant) contribution. 

The second difference with respect to the non-compact case is that the fluxes are restricted by quantization over internal cycles. If we write $z=L(x+i y)$, then
\be
\int_0^1 G_x \, dx= G^{(0)}_x =M \in 2\pi{\mathbb Z}\,,\qquad \int_0^1 G_y \, dy= G_y^{(0)} =N \in 2\pi{\mathbb Z},
\ee
where $G_1=G_x dx+ G_y dy$.
In holomorphic and anti-holomorphic components, this reads:
\be\label{eq:fluxquanta}
G_1^{(0)} =  \frac{1}{2L}\left(M-iN\right)dz+ \frac{1}{2L}\left(M+iN\right)d\bar{z}.
\ee

With this solution, the on-shell action is
\begin{eqnarray}
{\cal L}&=&-\Big|G_1^{(0)}-\frac{\bar{j}}{A_{T^2}}d\bar z\Big|^2
\nonumber\\
&=&-\left|\frac{j}{A_{T^2}}dz\right|^2 +\left[\left(G_1^{(0)},\frac{\bar{j}}{A_{T^2}}d\bar{z}\right)+{\text c.c.}\right]-\left|G_1^{(0)}\right|^2 \nonumber\\
&=& -2\,\frac{|j|^2}{A_{T^2}}+\frac{1}{\sqrt{A_{T^2}}}\left[j(M+iN)+{\text c.c.}\right]-(M^2+N^2).
\end{eqnarray}
As in the co-dimension one case, we obtain a finite contribution to the action, quadratic in the source $j$ (which again represents the bilinear gaugino condensate). We can set, if we wish, the discrete flux quanta to zero $N=M=0$, in which case the action is simply
\be
{\cal L}=-2\,\frac{|j|^2}{A_{T^2}}.
\ee
As we will see later, the dependence of these results on the transverse volume $A_{T^2}$ will serve as a non-trivial check on our methods.

\subsection{Gaugino condensation in D7-branes} \label{Sec:type_IIB}
We now address the case of the coupling between the bulk three-form flux of type IIB strings $G_3=F_3-\tau H_3$ and a gaugino bilinear $\lambda \lambda$ living on a localized D7-brane. The D7-brane wraps a holomorphic 4-cycle $\Sigma$ in a Calabi-Yau manifold $X$ with holomorphic (3,0)-form $\Omega_3$.\footnote{We normalize $\Omega_3$ by the volume of $X$ as $\int_X \Omega_3\wedge \ast\bar{\Omega}_3=V_{X}$.\label{foot:normalization}}

The relevant coupling was obtained in~\cite{Camara:2004jj}. Following the same steps as in~\eqref{eq:Gj}, it can be written as~\cite{Baumann:2010sx}:
\be
{\cal L}_{jG}=c \sqrt{g_s} \int_\Sigma \sqrt{g}\, G_3\cdot \Omega_3\, \bar\lambda\bar\lambda+{\text c.c.}=c \sqrt{g_s} \int_X  G_3\wedge \Omega_3\, (\bar\lambda\bar\lambda\,\delta^{(0)})+{\text c.c.}
\ee
Here we use the 10d Einstein frame and work in units with $\alpha'=1$. As in previous sections, we will only be interested in the structure of the effective action, and will be ignoring numerical factors, such as $c$, which we hence set to one. We will, however, pay attention to factors of $g_s$ and the internal volumes. These will provide a non-trivial check of our results, upon comparison with standard results of supergravity. 

As in~\cite{Camara:2004jj} we work in a factorized internal space $X=\Sigma\times T^2$, where the square torus is parametrized by a coordinate $z$. In this case, we can set $\Omega_3= dz\wedge \Omega_2$, where $\Omega_2$ is a holomorphic (2,0)-form on $\Sigma$. We also assume local tadpole cancellation so that the dilaton profile is constant, and we assume no warping and a vanishing $\tilde{F}_5$. Background fluxes will induce deviations which can be controlled in the large volume regime. We expect our results to generalize to more complicated setups. 

With these simplifications, we can directly apply the results of our previous toy models to the D7-brane case. We focus again on constant gaugino bilinears, i.e. on the gaugino zero modes relevant for the four-dimensional effective theory. As before, we expect the effective action for the three-form flux and gaugino coupling to take a perfect square form
\be\label{eq:actionD7}
{\cal L}\sim - g_s\left|G_3-P\left(\frac{\lambda\lambda}{\sqrt{g_s}}\delta^{(0)}\overline{\Omega}_3\right)\right|^2=-g_s\left|G_3-\frac{\lambda\lambda}{\sqrt{g_s}}\left(2d(\partial_z G_{T^2})+\frac{d\bar{z}}{A_{T^2}} \right)\wedge \overline{\Omega}_2 \right|^2,
\ee
where, as explained before, $P(\,)$ is a projector onto the subspace of closed forms.

The solution to the $G_3$ equations of motion is immediately obtained:
\be\label{eq:G3solution}
G_3= 2\frac{\lambda\lambda}{\sqrt{g_s}}d(\partial_z G_{T^2})\wedge \overline{\Omega}_2 + G^{(0)}_3,
\ee
where $G^{(0)}_3$ is a harmonic form. In a compact space flux quantization restricts the harmonic piece $G^{(0)}_3$, which is determined by integer flux quanta as in~\eqref{eq:fluxquanta}. Notice that the non-harmonic part contains a $(0,3)$ ISD component and a $(1,2)$ IASD one. The appearance of a $(1,2)$ component was studied in detail in~\cite{Baumann:2010sx}. 

Plugging the solution~\eqref{eq:G3solution} back into the action, we obtain both a quadratic and a quartic gaugino interaction in the four dimensional lagrangian
\begin{eqnarray}\label{eq:finiteaction}
{\cal L}=-g_s\left|G_3^{(0)}-\frac{\lambda\lambda}{\sqrt{g_s}A_{T^2}}\overline{\Omega}_3\right|^2=-|\lambda\lambda|^2\frac{V_{\Sigma}}{A_{T^2}}+\left[{\bar\lambda\bar\lambda}\frac{\sqrt{g_s}}{A_{T^2}}\left(G_3^{(0)},\overline{\Omega}_3\right)+\text{c.c.}\right]-g_s\left|G_3^{(0)}\right|^2
\end{eqnarray}
The derivation of these finite terms from the 10d action~\eqref{eq:actionD7}, and in particular the quartic gaugino interaction, is the main result of our paper.\footnote{ 
Strictly speaking, the no-scale structure of the potential will ensure that the last term in~\eqref{eq:finiteaction} will be canceled in the complete 4d action~\cite{Giddings:2001yu}. Our main interest in this paper, however, lies on the gaugino bilinear and quartic terms.} We will discuss their applications to moduli stabilization and construction of AdS vacua and dS uplift in future work. The structure of~\eqref{eq:finiteaction} will lead to a modifcation of the (trace-reversed) Einstein equations previously studied, so that the Maldacena-Nunez-like positivity arguments used in~\cite{Moritz:2017xto,Gautason:2018gln} would have to be reconsidered.

\subsubsection{Consistency with supersymmetry}
Our inclusion of a divergent quartic gaugino term in the action was motivated by similar discussions in Ho\v rava-Witten theory~\cite{Horava:1995qa,Horava:1996ma,Horava:1996vs,Mirabelli:1997aj}. The perfect square structure of the action obtained is reminiscent of heterotic strings~\cite{Dine:1985rz} where a 4d action of the (symbolic) form $|W_0+e^{-T}|^2$ was also shown to follow. As a further step, we had to project the gaugino source term onto the subspace of closed forms, so that the flux could compensate for it. The physically sensible finite result~\eqref{eq:finiteaction}, and the fact that the action can be zero in the non-compact case are good indications that we are on the right track.

In this subsection we perform a further check by comparing our result for the four-dimensional effective action to standard supergravity lagrangians. Our goal is to confirm that the four dimensional gaugino interactions in~\eqref{eq:finiteaction} have the correct dependence on K\"ahler moduli as determined by supersymmetry.\footnote{Although it should be possible, we are not keeping track on the precise dependence on complex structure moduli at this point.}

In order to do that we also need to consider the reduction from ten to four dimensions of the Einstein-Hilbert, as well as the gauge and gaugino kinetic terms. Up to numerical factors, dimensional reduction of the 10d Einstein frame action reads
\begin{eqnarray}
S&=&\int_{M^4\times X} d^{10}x\sqrt{-g}\,{\cal R}_{10}-\int_{M^4\times\Sigma}d^8x\sqrt{-g}\,\,\text{tr}\left(F_{mn}F^{mn}-i\lambda\slashed{D}\bar{\lambda}+c.c.\right)\nonumber\\
&=&A_{T^2}V_{\Sigma}\int_{M^4} d^{4}x\sqrt{-g_4}\,{\cal R}_{4}-{V_{\Sigma}}\int_{M^4}d^4x\sqrt{-g_4}\,\,\text{tr}\left(F_{\mu\nu}F^{\mu\nu}-i\lambda\slashed{D}\bar{\lambda}+c.c.\right),
\end{eqnarray}
where $m, n, ...$ are the indices for coordinates tangent to the D7-brane, and $\mu, \nu, ...$ for the uncompactified directions.

This action can be combined with the quartic and quadratic gaugino terms~\eqref{eq:finiteaction} obtained by integrating out $G_3$. In order to obtain the dependence on the K\"ahler moduli, we need to rescale the holomorphic three-form $\Omega_3\to \sqrt{A_{T^2}V_{\Sigma}}\,\Omega_3$, so that it is independent of K\"ahler moduli (c.f. footnote~\ref{foot:normalization}). Taking this into account, and rescaling the 4d metric as $g_{\mu\nu}\to (A_{T^2}V_{\Sigma})^{-1}g_{\mu\nu}$, we get
\begin{eqnarray}
{\cal L}\to \,{\cal R}_4 \,&-& {V_{\Sigma}}\,\text{tr}\left(F_{\mu\nu}F^{\mu\nu}-\frac{i}{(V_{\Sigma}A_{T^2})^{3/2}}\lambda\slashed{D}\bar{\lambda}+c.c.\right)\nonumber\\
&-&\frac{1}{V_{\Sigma}A_{T^2}^3}|\lambda\lambda|^2+\left[\frac{\sqrt{g_s}}{A_{T^2}^{5/2}V_{\Sigma}^{3/2}}{\bar\lambda\bar\lambda}\left(G_3^{(0)},\overline{\Omega}_3\right)+\text{c.c.}\right]
\end{eqnarray}
Finally, we bring these terms to their standard supersymmetry form by rescaling the gauginos as $\lambda\to (V_{\Sigma}A_{T^2})^{3/4}\lambda$:
\be
{\cal L}\to \, {\cal R}_4 \,- V_{\Sigma}\,\text{tr}\left(F_{\mu\nu}F^{\mu\nu}-i\lambda\slashed{D}\bar{\lambda}+\text{c.c.}\right)-{V_{\Sigma}^2}\,|\lambda\lambda|^2+\left[\frac{\sqrt{g_s}}{A_{T^2}}{\bar\lambda\bar\lambda}\left(G_3^{(0)},\overline{\Omega}_3\right)+\text{c.c.}\right].
\ee
This result is remarkable. The dependence on the K\"ahler moduli $V_\Sigma  \sim g_{\rm YM}^{-2}$, $A_{T^2}$ and on the dilaton $S\sim g_s^{-1}$ of the gaugino bilinear and quartic operators are precisely the ones required by supersymmetry (see, e.g.~\cite{Wess:1992cp}, Appendix G).\footnote{This can be explicitly checked in the case in which $\Sigma$ is a factorizable torus $\Sigma=T^2\times T^2$. In this case the standard type IIB K\"ahler potential is
\be
K\sim -\sum_{i=1}^3\log(T_i+T_i^*)-\log(S+S^*)\sim -2\log(8V_\Sigma A_{T^2})+\log(2g_s)\,,
\ee
the gauge kinetic function is $f \sim T_1\sim V_\Sigma$, and the superpotential $W\sim(G_3^{(0)},\overline{\Omega}_3)$ is independent of the K\"ahler moduli (this represents the Gukov-Vafa-Witten superpotential~\cite{Gukov:1999ya}).}

As in~\eqref{eq:separated}, we could add different sets of D7 branes separated in the internal space. Integrating  out the three-form bulk flux would result in quartic fermionic interactions between distant gaugino bilinears. These terms are also expected from supersymmetry. Our methods clearly show how these finite contributions arise in a controlled manner.  We regard the results of this subsection as a highly non-trivial check on our proposed action~\eqref{eq:actionD7}.

\section{Conclusions}\label{Sec:conclusion}
In this paper, we have analyzed in detail the interactions between bulk fluxes and gaugino bilinears localized on D7-branes. The flux sourced by the gauginos diverges on the brane locus, which leads naively to a divergent action. Inspired by the resolution of analogous divergences in Ho\v rava-Witten theory, we have proposed a quartic gaugino counterterm that makes the action manifestly finite. The resulting action takes a perfect square structure reminiscent of heterotic and Ho\v rava-Witten theories.  

In the non-compact case, the background $G_3$ flux can always compensate a non-vanishing gaugino bilinear, such that the action vanishes. We have furthermore studied the effect of compactification and flux quantization on the resulting 4d effective action. In compact setups flux quantization prevents the action from vanishing and a finite contribution involving the gaugino bilinear remains. In the presence of non-zero quantized fluxes, this includes a gaugino mass term.

Our proposal passes the highly non-trivial test of reproducing the gaugino bilinear and quartic terms in 4d supergravity. To our knowledge, this consistency check is not met by previous attempts of writing down a 10d action describing the brane gaugino condensate in the presence of bulk fluxes. Regardless of the application of our results to dS uplift, we believe that the divergence-free 10d action proposed in this paper is the correct starting point for studying flux compactifications with
brane gauginos from a 10d perspective in general.

Hopefully, one application of this work will be in the study of gaugino condensation in type IIB theories. We expect our results to have implications in the understanding of moduli stabilization and the search for AdS vacua and dS uplift from a ten dimensional perspective. In particular, recent concerns about the KKLT proposal based on Maldacena-Nunez-type no-go argument would need to be reconsidered in light of our results.
The 10d action we propose will lead to a drastic modification of the energy momentum tensor, which may change the conclusions of recent studies. We will investigate this issue in future work. 

\section*{Acknowledgments}
We would like to thank Thomas Van Riet, Lukas Witkowski and Fengjun Xu for useful discussions. The work of YH is supported by the Advanced ERC grant SM-grav, No 669288. YH thanks the hospitality of the AstroParticule et Cosmologie (APC). The work of GS is supported in part by the DOE grant DE-SC0017647 and the Kellett Award of the University of Wisconsin.
We would like to thank the hospitality of Simons Center for Geometry and Physics, where this work was initiated during the 2018 Simons Summer workshop.

\bibliography{Bibliography}\bibliographystyle{utphys}

\end{document}